\documentclass{article}
\usepackage{spconf,amsmath,graphicx}
\usepackage{siunitx, booktabs, tabularx, graphicx, multirow, subcaption, adjustbox}
\usepackage{fancyhdr}

\title{Evaluating Identity Leakage in Speaker De-Identification Systems}
\name{Seungmin Seo, Oleg Aulov, Afzal Godil, Kevin Mangold}

\address{National Institute of Standards and Technology, Gaithersburg, MD, USA}
\fancypagestyle{firstpage}{
  \fancyhf{} 
  \fancyfoot[C]{\small This work has been submitted to the IEEE for possible publication. 
  Copyright may be transferred without notice, after which this version may no longer be accessible.}
}

\begin{document}
%
\thispagestyle{firstpage}
\maketitle

\begin{abstract}
Speaker de-identification aims to conceal a speaker’s identity while preserving intelligibility of the underlying speech.  We introduce a benchmark that quantifies residual identity leakage with three complementary error rates: equal error rate, cumulative match characteristic hit rate, and embedding-space similarity measured via canonical correlation analysis and Procrustes analysis. Evaluation results reveal that all state-of-the-art speaker de-identification systems leak identity information. The highest performing system in our evaluation performs only slightly better than random guessing, while the lowest performing system achieves a 45\% hit rate within the top 50 candidates based on CMC. These findings highlight persistent privacy risks in current speaker de-identification technologies.

\end{abstract}
\begin{keywords}
speaker de-identification, voice privacy, identity leakage
\end{keywords}
\section{Introduction}
\label{sec:intro}

The speech we stream through videoconferencing platforms, voice assistants, and call-center recorders conveys far more than lexical content: it embeds biometric signatures that can single out an individual. Recent privacy statutes—most prominently the EU’s General Data Protection Regulation (GDPR)—explicitly classify these signatures as personally identifiable information \cite{GDPR2016}.

Consequently, speaker de-identification (SDID) systems that operate on live, spontaneous speech have become a research priority. Unlike offline voice-conversion or text-to-speech pipelines, real-time SDID must satisfy millisecond-scale latency budgets and preserve intelligibility and naturalness, while withstanding attacks from state-of-the-art speaker-recognition models \cite{VoicePrivacy2020}.

Individual components—e.g.\ disentangled speaker–content representation learning \cite{Qian2019AutoVC} and neural audio codecs \cite{Zeghidour2021Soundstream} — have shown promise, yet the field still lacks a rigorous answer to a central question: \textbf{How much identity information “leaks” through today’s end-to-end SDID pipelines?}

Prior studies are difficult to compare \cite{Li2025SpecWavAttack, Li2025HLTCOE, Mawalim2025TitaNetLarge, Zhang2025AugmentedFeatureAttack}; most rely on a single speaker–recognition back-end, and a solitary metric such as equal error rate (EER). To advance beyond this fragmented landscape, we introduce a \textbf{multi-view identity-leakage evaluation suite} that integrates EER, cumulative match characteristic (CMC) analysis, and embedding-space similarity measured with canonical correlation analysis (CCA) followed by Procrustes alignment.  

Each perspective exposes a distinct facet of residual speaker information: EER quantifies binary verification risk, CMC reflects search-rank leakage, and the embedding analysis localises where representations converge in latent space. Each SDID system was required to meet the real-time processing budget, evaluated independently by the other test-and-evaluation agency; the present paper concentrates on privacy metrics. Under this protocol, every system leaks identity: the best performance achieved exceeds random guessing only marginally yet still significantly, whereas the weakest reaches a 45\% hit-rate among the top-50 candidates on CMC. These findings underscore the persistent challenge of robust, privacy-preserving speaker de-identification.

\section{Speaker De-identification Systems}\label{sec:sdid}

The five SDID systems in this study were submitted to NIST for evaluation—all developed under the IARPA ARTS program\footnote{www.iarpa.gov/research-programs/arts}—including four performer systems and one baseline built by a Test \& Evaluation partner. Note that no system descriptions are publicly available at the time of this writing, so the references reflect relevant work by the same researchers. \cite{wang2025diffattack, wang2024toward,xinyuan2024hltcoe}

\begin{table*}[t]
\centering
\caption{Summary of SDID systems, methods, and training datasets.}
\label{tab:full_system_summary}
\small
\renewcommand{\arraystretch}{1.15}
\setlength{\tabcolsep}{5pt}
\begin{tabular}{p{2cm} p{4cm} p{5.3cm} p{5cm}}
\toprule
\textbf{System} & \textbf{Method} & \textbf{Key Components} & \textbf{Training Data} \\
\midrule
VOXLET \newline(Galois) & Latent perturbation + vocoder & wav2vec 2.0 encoder, HiFiGAN 2.0 vocoder, DP noise in latent space & Encoder: LibriSpeech, LibriVox, TIMIT; Vocoder: DAPS + MIT IRs + REVERB/ACE \\
\midrule
RASP \newline(Honeywell) & Disentangled autoencoder & HuBERT (content); speaker replaced with pseudo-embedding (ControlVCSpawn); pitch/energy via variance encoder & HuBERT: LibriSpeech; Speaker encoder: VoxCeleb1 + LibriSpeech; Full model: VCTK \\
\midrule
SHADOW (JHU) & Token conversion + speaker swap & EnCodec + Wav2Vec2 tokens; PLDA for speaker swapping; FreeVC module & Main: Libri-Light, MLS, LibriTTS-R; FreeVC: VCTK; PLDA: VoxCeleb2, LibriTTS-R \\
\midrule
PHORTRESS (SRI) & Articulatory synthesis - SPARC & Source extractor, EMA inverter, speaker encoder, DDSP vocoder, fabricated speaker embedding & GigaSpeech, LibriTTS-R, VCTK; Speaker embedding: Pyannote \\
\midrule
kNN-VC (LLNL-baseline) & $k$-NN in WavLM space & Frame-wise regression using nearest neighbors; re-synthesized via HiFiGAN; fusion across pseudo-profiles & Latent space: WavLM; Vocoder: HiFiGAN trained on LibriSpeech \\
\bottomrule
\end{tabular}
\end{table*}

Each system takes as an input a streaming speech segment and outputs a streaming modified version designed to conceal the speaker's identity. The primary goals are (1) to prevent speaker-recognition models from linking original and de-identified segments, and (2) to ensure that de-identified segments generated for the same speaker (under the same or different anonymization profiles) are either consistent or distinct as appropriate.

\section{Evaluation}
\subsection{Data}
The evaluation set is derived from the Mixer 3 corpus \cite{Cieri2007Mixer}. We retained only native American English speakers with at least five recording sessions, yielding 223 speakers in total; 180 of these have ten or more sessions. For each session, the first 60 seconds were discarded to remove greetings, call-setup noise, and channel-stabilization artifacts, following common practice in speaker-recognition evaluations. The remaining audio was segmented, in round-robin order, into non-overlapping chunks containing at least 10, 30, and 60 seconds of detected speech activity, as determined by the LDC Broad Phonetic Class SAD \cite{ldc-bpcsad}.

\begin{table}[th]
    \centering
    \caption{Evaluation data statistics. (M: male, F: female)} 
    \vspace*{0.5\baselineskip}
    \label{tab:trial_segment} 
    \resizebox{\linewidth}{!}{%
        \begin{tabular}{cccc}
        \#Speakers (M/F) & \#Segments & \#Target trials & \#Non-target trials \\
        \hline
        223 (81/142) & 2,983 & 10,458 & 940,567 \\
        \hline
        \end{tabular}
    }
\end{table}

\begin{table}[th]
    \centering
    \caption{Number of unique speakers by age group and sex.}
    \vspace*{0.5\baselineskip}
    \label{tab:speaker_agegroup}
    \resizebox{0.5\linewidth}{!}{%
        \begin{tabular}{ccc}
        Age Group & Male & Female \\
        \hline
        Young (17--24) & 17 & 15 \\
        \hline
        Adult (25--54) & 57 & 104 \\
        \hline
        Senior (55--85) & 7 & 23 \\
        \hline
        \end{tabular}
    }
\end{table}

\subsubsection{Trials}

The trial design assesses the performance of the SDID system under various conditions. Relevant data statistics are presented in Tables \ref{tab:trial_segment} and \ref{tab:speaker_agegroup}. The trial lists are categorized into scenario-specific sets, as outlined in Table \ref{tab:trial_scenarios_part1}. Trial set 1 consists of target trials comparing original speech against anonymized speech segments of the same speaker, and non-target trials comparing original speech segments of one speaker with anonymized segments of different speakers. Trial set 2 consists solely of anonymized segments, where target trials consist of different segments of the same anonymous voice. In the evaluation, the SDID system's ability to generate multiple distinct consistent pseudo-voices per speaker was assessed. Each speaker had eight pseudo-voices generated using different anonymization profiles; two of these were selected for testing. For each pseudo-voice, multiple segments from different sessions were used to test consistency. The cross-profile trial sets evaluate whether pseudo-voices from the same speaker remain acoustically distinct from one another and from pseudo-voices of other speakers. For each set, the expected EER is reported.

\begin{table}[t]
    \centering
    \caption{Different trial scenarios for evaluating SDID performance - \textbf{orig}: original speech segment of a given speaker, \textbf{deID}: de-identified speech segment of a given speaker with a single pseudo-profile, \textbf{deID {\boldmath$P_{n}$}}: de-identified speech segment of a given speaker with pseudo-profile $n$ generated by the SDID.} 
    \vspace*{0.5\baselineskip}
    \label{tab:trial_scenarios_part1} 
    \resizebox{\linewidth}{!}{%
        \begin{tabular}{llll}
        \hline
        & Target & Non-target & EER expectation \\
        \hline
        Trial set 1 & orig - deID & orig - deID & close to 50\% \\
        \hline
        Trial set 2 & deID - deID & deID - deID & close to 0\% \\
        \hline
        Cross-profile & deID $P_{1}$ - deID $P_{1}$ & deID $P_{1}$ - deID $P_{2}$ & close to 0\% \\
        \hline
        \end{tabular}
    }
\end{table}

\subsection{Speaker Identification Systems}
To evaluate the performance of SDID systems, we employed three speaker identification (SID) models, each built on distinct architectures and trained with different strategies.

NVIDIA NeMo’s TitaNet-L is built on 1D separable convolutions, SE layers, and a ContextNet encoder~\cite{koluguri2022titanet}, trained end-to-end with angular softmax loss. It maps speech to fixed-length embeddings (tvectors) and was trained on VoxCeleb 1/2, RIR noise, Fisher, Switchboard, and LibriSpeech.

ECAPA-TDNN~\cite{dawalatabad2104ecapa}, also from NVIDIA NeMo Framework, enhances TDNNs with Res2Blocks, SE layers, and a frequency-channel attention module tailored for short speech. Its dual-path design captures both local and temporal cues and uses the same training data as TitaNet-L.

SRI’s OLIVE 5.3.1 Martini plugin takes a different approach, using a TDNN-based xvector extractor trained on PNCC features and a PLDA backend~\cite{mclaren2020adaptive}. It includes a DNN-based speech activity detector and was trained on NIST SRE 2004–2012, Mixer 6, and VoxCeleb 1/2.

\begin{table*}[t]
  \centering
  \footnotesize
  \caption{Side-by-side comparison of trial set 1 (“orig vs de-ID”) and trial set 2 
           (“identity consistency / collision”) – EER (\%).
           $P_{n}$ denotes the trial sets consist of de-identified segments for pseudo-profile $n$.}
  \label{tab:eer_result}
  \resizebox{\linewidth}{!}{%
    \begin{tabular}{lcccccc|cccccc}
      \multirow{3}{*}{SDID system} 
          & \multicolumn{6}{c|}{\textbf{Trial-1: Original vs De-ID}} 
          & \multicolumn{6}{c}{\textbf{Trial-2: Consistency / Collision}} \\
      \cline{2-13}
          & \multicolumn{2}{c}{TitaNet-L} 
          & \multicolumn{2}{c}{ECAPA-TDNN} 
          & \multicolumn{2}{c|}{Olive}
          & \multicolumn{2}{c}{TitaNet-L} 
          & \multicolumn{2}{c}{ECAPA-TDNN} 
          & \multicolumn{2}{c}{Olive} \\
      \cline{2-13}
          & $P_{1}$ & $P_{2}$ & $P_{1}$ & $P_{2}$ & $P_{1}$ & $P_{2}$ 
          & $P_{1}$ & $P_{2}$ & $P_{1}$ & $P_{2}$ & $P_{1}$ & $P_{2}$ \\
      \hline
      \textit{EER w/o SDID} 
          & \multicolumn{6}{c|}{3.75 / 3.59 / 4.39}  
          & \multicolumn{6}{c}{—} \\[-0.15em]
      Baseline   & 34.17 & 34.78 & 35.82 & 38.76 & 34.05 & 32.24
                 & 1.14  & 1.42  & 1.28  & 1.37  & 1.09  & 1.38  \\
      VOXLET     & 33.73 & 33.58 & 34.18 & 33.58 & 36.57 & 36.06
                 & 42.39 & 42.49 & 44.29 & 43.34 & 41.95 & 42.32 \\
      RASP       & 33.01 & 37.45 & 38.02 & 41.82 & 29.97 & 34.03
                 & 6.10  & 6.01  & 9.46  & 9.09  & 5.82  & 12.43 \\
      SHADOW     & 42.63 & 43.44 & 46.03 & 44.77 & 43.25 & 43.10
                 & 26.06 & 26.49 & 26.47 & 26.83 & 21.94 & 26.78 \\
      PHORTRESS  & 50.03 & 49.12 & 51.14 & 47.86 & 46.57 & 46.78
                 & 19.23 & 19.02 & 22.89 & 21.02 & 24.90 & 24.15 \\
      \hline
    \end{tabular}
  }
\end{table*}

\subsection{Anonymization Effectiveness}
Trial set 1 evaluates how well SDID systems break the link between original and de-identified utterances. Target trials pair "Alice" with "pseudo-Alice"; non-targets pair "Alice" with "pseudo-Bob." An EER near 50\% reflects strong privacy. Without anonymization, the SID systems achieve (3.6–4.4)\% EER, confirming their high baseline accuracy. After anonymization, EERs rise substantially to (29–51)\% depending on the system and profile (left half of Table \ref{tab:eer_result}). EER differences across SID systems and between $P_{1}$/$P_{2}$ stay within a few points, suggesting architecture- and sample-agnostic gains. While SDID systems degrade recognizability, most do not fully obscure identity - as evidenced by EER below 50\% — suggesting that perfect anonymity remains out of reach.

\subsection{Anonymization Stability and Profile Collisions}

Trial set 2 evaluates whether SDID systems maintain consistent pseudo-voices across utterances. The baseline system, which reuses a single, fixed synthetic voice per speaker, achieves an EER of $\approx$1\%, indicating high within-profile consistency and minimal confusion with other pseudo-voices.

In contrast, most SDID systems yield (20–44)\% EERs (Table \ref{tab:eer_result}, right), suggesting pseudo-identities are often acoustically inconsistent and confusable. This drift undermines identity separation and reveals a secondary vulnerability: instability in voice characteristics may expose the synthetic origin of the speech, even when de-identification prevents direct speaker recovery.

\subsection{Same-speaker Anonymization Profile Distinctness}

Cross-profile testing evaluates whether anonymized segments generated from different pseudo profiles for the same speaker can be reliably distinguished. Each trial asks a verifier to determine whether two segments originate from the same pseudo-profile (e.g., “pseudo-Alice 1” vs.\ “pseudo-Alice 1”) or from two different ones (e.g., “pseudo-Alice 1” vs.\ “pseudo-Alice 2”). If the SDID system produces acoustically distinct, person-like pseudo-voices, the expected EER should approach zero. Higher EERs indicate overlap between the acoustic spaces of different pseudo-profiles, enabling downstream models—or potential attackers—to link them. Across 10,458 target and 20,916 non-target trials, Table~\ref{tab:merged-eval} shows that EERs range from 8\% to 51\%. The RASP system achieves low EERs, indicating that its pseudo-voices are largely separable. In contrast, systems like VOXLET yield EERs near 50\%, suggesting that pseudo-profiles frequently overlap and are difficult to distinguish. This overlap is not benign: it enables clustering of de-identified utterances and reveals that the speech is synthetic, introducing a new vector for identity and privacy leakage.

\begin{table}[t]
  \centering
  \caption{System-level evaluation: AUC-CMC, mean identification rank, and cross-profile EER (\%).}
  \label{tab:merged-eval}
  \vspace*{0.5\baselineskip}
  \resizebox{\linewidth}{!}{%
    \begin{tabular}{lcccccc}
      \textbf{Metric} & \textbf{Random} & \textbf{Baseline} & \textbf{VOXLET} & \textbf{RASP} & \textbf{SHADOW} & \textbf{PHORTRESS} \\
      \midrule
      AUC-CMC         & 0.015 & \textbf{0.169} & \textbf{0.373} & \textbf{0.175} & \textbf{0.072} & \textbf{0.067} \\
      Mean Rank       & 736   & \textbf{274}   & \textbf{158}   & \textbf{273}   & \textbf{547}   & \textbf{540}   \\
      TitaNet-L EER   & –     & 1.2            & 50.6           & 8.2            & 33.6           & 35.5           \\
      ECAPA-TDNN EER  & –     & 1.4            & 49.9           & 11.3           & 33.8           & 36.2           \\
      \bottomrule
    \end{tabular}
}
\end{table}


\subsection{Measuring Identity Leakage}
\label{sec:identity-leakage}

\begin{table}[t]
  \centering
  \caption{CMC hit-rate (\%) at \textit{k}.
           Results for each system are averaged over two pseudo-profiles and the two SID models.}
  \vspace*{0.5\baselineskip}
  \label{tab:cmc-hit-rate}
  \sisetup{group-minimum-digits=4,table-number-alignment=center} 
  \resizebox{\linewidth}{!}{%
        \begin{tabular}{
            l                                                    
            S[table-format=2.2]                                   
            S[table-format=2.2]                                   
            S[table-format=2.2]                                   
            S[table-format=2.2]                                   
            S[table-format=2.2]                                   
            S[table-format=2.2]                                   
        }
        \multicolumn{1}{c}{Rank-$k$} &
        \multicolumn{1}{c}{Random} &
        \multicolumn{1}{c}{Baseline} &
        \multicolumn{1}{c}{VOXLET} &
        \multicolumn{1}{c}{RASP} &
        \multicolumn{1}{c}{SHADOW} &
        \multicolumn{1}{c}{PHORTRESS} \\
        \midrule
        1  & 0.48 & \textbf{3.01} & \textbf{12.29} & \textbf{3.02} & \textbf{0.86} & \textbf{0.72} \\
        5  & 2.39 & \textbf{9.25} & \textbf{25.60} & \textbf{8.41} & \textbf{3.08} & \textbf{2.73} \\
        10 & 4.73 & \textbf{12.32} & \textbf{31.15} & \textbf{11.27} & 4.29 & 4.07 \\
        20 & 9.22 & \textbf{16.63} & \textbf{37.52} & \textbf{15.48} & 6.59 & 6.20 \\
        50 & 21.48 & \textbf{24.80} & \textbf{45.05} & \textbf{24.89} & 11.39 & 10.53 \\
        \bottomrule
        \end{tabular}
  }    
  \vspace*{-0.5\baselineskip}
\end{table}


\begin{table}[t]
  \centering
  \caption{Embedding-space similarity across SDID systems, measured with CCA and Procrustes alignment. \textbf{CCA-Mean$_{\text{Top10}}$}: mean of the ten largest canonical correlations between the two embedding sets; \textbf{P-MSE}: Procrustes mean-squared error after optimal orthogonal alignment; \textbf{P-COSINE}: average cosine similarity after the same alignment.}
  \vspace*{0.5\baselineskip}
  \label{tab:cca-proc}
  \resizebox{\linewidth}{!}{%
  \sisetup{round-mode=places, round-precision=3,
           table-number-alignment=center}
      \begin{tabular}{
          l                                          
          S[table-format=1.3]                        
          S[table-format=1.3]                        
          S[table-format=1.3]                        
          S[table-format=1.3]                        
          S[table-format=1.3]                        %
          S[table-format=1.3]                        
      }
        Metric & {Random} & {Baseline} & {VOXLET} & {RASP} & {SHADOW} & {PHORTRESS} \\
        \midrule
        CCA-Mean$_{\text{Top10}}$      & 0.459 & \textbf{0.875} & \textbf{0.627} & \textbf{0.854} & \textbf{0.597} & \textbf{0.760} \\
        P-MSE     & 0.008 & \textbf{0.005} & \textbf{0.007} & \textbf{0.005} & \textbf{0.007} & \textbf{0.006} \\
        P-COSINE     & 0.212 & \textbf{0.502} & \textbf{0.330} & \textbf{0.488} & \textbf{0.316} & \textbf{0.394} \\
        \bottomrule
      \end{tabular}%
  }
\end{table}

For each original and de-identified speech segment, we extract \textit{x-vectors} using two SID models — TitaNet-L and ECAPA-TDNN — and compare the resulting embeddings. We report three metrics:

\begin{enumerate}
  \item CMC hit rates at selected ranks: the proportion of queries where the correct identity appears within the top-$k$ nearest neighbors (Table~\ref{tab:cmc-hit-rate}). CMC@k is the rate at which a speaker is correctly identified among the top $k$ candidates.
  \item Area under the CMC curve (AUC-CMC) and the corresponding mean identification rank (Table~\ref{tab:merged-eval}), which summarize the full CMC curve. Higher AUC or lower mean rank indicates weaker privacy; values near the random baseline indicate stronger de-identification.
  \item Embedding-space similarity, measured via:(i) mean CCA between the top ten singular vectors of the original and anonymized x-vector spaces, and (ii) Procrustes analysis (MSE and cosine similarity) after orthogonal alignment (Table~\ref{tab:cca-proc}).
\end{enumerate}

\subsubsection{CMC analysis}
As expected, the random baseline—obtained by shuffling speaker labels—exhibits minimal identity retention, with a rank-1 hit rate of just 0.48\%.  
All real systems leak identity to to varying degrees.  
VOXLET is the most permissive, achieving a rank-1 hit rate of 12.29\% and a rank-50 hit rate of 45.05\%.  
The Baseline and RASP systems perform similarly (rank-1 $\approx$3\%, rank-50 $\approx$25\%), while SHADOW and PHORTRESS reduce leakage to below 1\% at rank-1 and below 11\% at rank-50.  
These trends are reflected in the AUC-CMC figures of Table~\ref{tab:merged-eval}: VOXLET (0.373) retains more than twice the identity information of the next most leaky systems (Baseline 0.169, RASP 0.175), while SHADOW and PHORTRESS reduce AUC-CMC by over 50\%.

\subsubsection{Embedding-space similarity}
CCA and Procrustes metrics offer an orthogonal view that is less sensitive to gallery size. CCA answers the question: \textbf{“Does a linear projection exist in which the de-identified vectors still line up with the originals?”}  When the leading canonical correlations exceed about \(0.6\), we interpret this as strong linear dependence and thus potential attribute leakage; values near zero imply that the SDID system has largely destroyed linear cues.

Procrustes analysis addresses a complementary question: \textbf{“Are the two embedding clouds almost identical after some rigid rotation or reflection?”} After fitting the optimal rotation we report MSE and the mean cosine similarity between aligned pairs; larger errors or lower cosines indicate that the global manifold has been effectively scrambled.

Interestingly, Baseline and RASP achieve the strongest alignment (\(\text{CCA-Mean}_{\text{Top10}}=0.875\) and \(0.854\)) despite leaking less identity in the CMC than VOXLET. Conversely, VOXLET shows the highest leakage but only a moderate CCA score (\(0.627\)), suggesting its privacy weakness stems from fine-grained nearest-neighbor structure rather than global sub-space overlap. Procrustes results reinforce this view: Baseline and RASP yield the lowest reconstruction error (MSE \(0.005\)) and highest cosine similarity (\(0.502\) and \(0.488\)), whereas VOXLET is closer to SHADOW and PHORTRESS.

\section{Conclusion}
\label{sec:con}
The multi-view analysis shows that identity leakage is universal but heterogeneous. Baseline and RASP leak far less at rank-1 (about \(3\%\)) yet exhibit the strongest global alignment with the original space (\(\text{CCA}_{\text{Top10}}=0.875\) and \(0.854\); Procrustes MSE \(0.005\)), suggesting that their residual risk stems from broad sub-space similarity. By contrast, SHADOW and PHORTRESS suppress neighborhood retrieval almost to chance (rank-1 \(<1\%\), rank-50 \(\le 11\%\)) but leave moderate sub-space overlap (\(\text{CCA}_{\text{Top10}}\approx(0.60\text{--}0.76)\)).

This decoupling confirms that removing local identity cues and rotating the global manifold are orthogonal privacy levers; optimizing one does not guarantee the other. Consequently, single-metric evaluations can misrepresent risk.

Every SDID system left detectable traces: using our metric, the highest scoring system exceeded random guessing by a small but significant margin, while the lowest one still reached a 45\% rank-50 hit-rate. Leakage patterns varied—some systems suppressed rank-1 retrieval yet preserved global embedding alignment, others did the opposite—showing that single metrics are insufficient for privacy claims.

\section{Acknowledgments and Disclaimers}
\label{sec:ack}

This research is based upon work supported by the Office of the Director of National Intelligence (ODNI), Intelligence Advanced Research Projects Activity (IARPA), Anonymous Real-Time Speech (ARTS) research program, under Interagency Agreement (IAA) IARPA-20001-D250300042 with NIST.

Certain equipment, instruments, software, or materials are identified in this paper in order to specify the experimental procedure adequately. Such identification is not intended to imply recommendation or endorsement of any product or service by NIST, nor is it intended to imply that the materials or equipment identified are necessarily the best available for the purpose. Opinions, views, recommendations, findings, and conclusions contained herein are those of the authors and should not be interpreted as necessarily representing the official views, policies or endorsements, either expressed or implied, of the ODNI, IARPA, NIST or the U.S. Government.


\bibliographystyle{IEEEbib}
\bibliography{manuscript}

\begin{thebibliography}{10}

\bibitem{GDPR2016}
{European Parliament and Council},
\newblock ``Regulation (eu) 2016/679: General data protection regulation,'' 2016,
\newblock Official Journal of the European Union.

\bibitem{VoicePrivacy2020}
Julian R.~F. Fang, Tomi Kinnunen, and Patrick~L. et~al.,
\newblock ``The voiceprivacy 2020 challenge: {Speaker} anonymisation for privacy preservation,''
\newblock in {\em Proc. \textit{INTERSPEECH}}, 2020.

\bibitem{Qian2019AutoVC}
Kaizhi Qian, Yang Zhang, Shiyu Chang, Xuesong Yang, and Mark Hasegawa-Johnson,
\newblock ``Autovc: Zero-shot voice style transfer with only autoencoder loss,''
\newblock in {\em International Conference on Machine Learning}. PMLR, 2019, pp. 5210--5219.

\bibitem{Zeghidour2021Soundstream}
Neil Zeghidour, Alejandro Luebs, Ahmed Omran, Jan Skoglund, and Marco Tagliasacchi,
\newblock ``Soundstream: An end-to-end neural audio codec,''
\newblock 2021, vol.~30, pp. 495--507, IEEE.

\bibitem{Li2025SpecWavAttack}
Yuqi Li, Yuanzhong Zheng, Zhongtian Guo, Yaoxuan Wang, Jianjun Yin, and Haojun Fei,
\newblock ``Specwav-attack: Leveraging spectrogram resizing and wav2vec 2.0 for attacking anonymized speech,''
\newblock in {\em ICASSP 2025-2025 IEEE International Conference on Acoustics, Speech and Signal Processing (ICASSP)}. IEEE, 2025, pp. 1--2.

\bibitem{Li2025HLTCOE}
Henry~Li Xinyuan, Ashi Garg, Zexin Cai, Kevin Duh, Leibny~Paola Garc{\'\i}a-Perera, Sanjeev Khudanpur, Nicholas Andrews, and Matthew Wiesner,
\newblock ``Hltcoe submission to the voiceprivacy attacker challenge,''
\newblock in {\em ICASSP 2025-2025 IEEE International Conference on Acoustics, Speech and Signal Processing (ICASSP)}. IEEE, 2025, pp. 1--2.

\bibitem{Mawalim2025TitaNetLarge}
Candy~Olivia Mawalim, Aulia Adila, and Masashi Unoki,
\newblock ``Fine-tuning titanet-large model for speaker anonymization attacker systems,''
\newblock in {\em ICASSP 2025-2025 IEEE International Conference on Acoustics, Speech and Signal Processing (ICASSP)}. IEEE, 2025, pp. 1--2.

\bibitem{Zhang2025AugmentedFeatureAttack}
Yanzhe Zhang, Zhonghao Bi, Feiyang Xiao, Xuefeng Yang, Qiaoxi Zhu, and Jian Guan,
\newblock ``Attacking voice anonymization systems with augmented feature and speaker identity difference,''
\newblock in {\em ICASSP 2025-2025 IEEE International Conference on Acoustics, Speech and Signal Processing (ICASSP)}. IEEE, 2025, pp. 1--2.

\bibitem{wang2025diffattack}
Qing Wang, Jixun Yao, Zhaokai Sun, Pengcheng Guo, Lei Xie, and John~HL Hansen,
\newblock ``Diffattack: Diffusion-based timbre-reserved adversarial attack in speaker identification,''
\newblock in {\em ICASSP 2025-2025 IEEE International Conference on Acoustics, Speech and Signal Processing (ICASSP)}. IEEE, 2025, pp. 1--5.

\bibitem{wang2024toward}
Zhenyu Wang and John~HL Hansen,
\newblock ``Toward improving synthetic audio spoofing detection robustness via meta-learning and disentangled training with adversarial examples,''
\newblock {\em IEEE Access}, vol. 12, pp. 99894--99911, 2024.

\bibitem{xinyuan2024hltcoe}
Henry~Li Xinyuan, Zexin Cai, Ashi Garg, Kevin Duh, Leibny~Paola Garc{\'\i}a-Perera, Sanjeev Khudanpur, Nicholas Andrews, and Matthew Wiesner,
\newblock ``Hltcoe jhu submission to the voice privacy challenge 2024,''
\newblock {\em arXiv preprint arXiv:2409.08913}, 2024.

\bibitem{Cieri2007Mixer}
Christopher Cieri and David Miller,
\newblock ``Mixer 3 speech audio collection,''
\newblock in {\em Proc. \textit{LREC}}, 2007.

\bibitem{ldc-bpcsad}
Neville Ryant,
\newblock ``{L}inguistic {D}ata {C}onsortium {B}road {P}honetic {C}lass {S}peech {A}ctivity {D}etector (ldc-bpcsad),'' 2023.

\bibitem{koluguri2022titanet}
Nithin~Rao Koluguri, Taejin Park, and Boris Ginsburg,
\newblock ``Titanet: Neural model for speaker representation with 1d depth-wise separable convolutions and global context,''
\newblock in {\em ICASSP 2022-2022 IEEE International Conference on Acoustics, Speech and Signal Processing (ICASSP)}. IEEE, 2022, pp. 8102--8106.

\bibitem{dawalatabad2104ecapa}
Nauman Dawalatabad, Mirco Ravanelli, Fran{\c{c}}ois Grondin, Jenthe Thienpondt, Brecht Desplanques, and Hwidong Na,
\newblock ``Ecapa-tdnn embeddings for speaker diarization,''
\newblock in {\em Proc. Interspeech 2021}, 2021, pp. 3560--3564.

\bibitem{mclaren2020adaptive}
Mitchell McLaren, Md~Hafizur Rahman, Diego Cast{\'a}n, Mahesh~Kumar Nandwana, and Aaron Lawson,
\newblock ``Adaptive mean normalization for unsupervised adaptation of speaker embeddings.,''
\newblock in {\em Odyssey}, 2020, pp. 88--94.

\end{thebibliography}

\end{document}